# Hot Electron Dynamics in Plasmonic Thermionic Emitters

Nicki Hogan[1], Shengxiang Wu[1], Matthew Sheldon*[1,2]

[1]Department of Chemistry, Texas A&M University, College Station, TX, 77843-3255, USA.

[2]Department of Material Science and Engineering, Texas A&M University, College Station, TX, 77843-3255, USA.

*sheldonm@tamu.edu

**Abstract:**

Thermionic converters generate electricity from thermal energy in a power cycle based on vacuum emission of electrons. While thermodynamically efficient, practical implementations are limited by the extreme temperatures required for electron emission (> 1500 K). Here, we show how metal nanostructures that support resonant plasmonic absorption enable an alternative strategy. The high temperatures required for efficient vacuum emission can be maintained in a sub-population of photoexcited "hot" electrons during steady-state optical illumination, while the lattice temperature of the metal remains within the range of thermal stability, below 600 K. We have also developed an optical thermometry technique based on anti-Stokes Raman spectroscopy that confirms these unique electron dynamics. Thermionic devices constructed from optimized plasmonic absorbers show performance that can out-compete other strategies of concentrated solar power conversion in terms of efficiency and thermal stability.

**Main Text:**

Metals that support surface plasmons, the coherent oscillation of free electrons, provide efficient coupling of electromagnetic radiation into extremely confined sub-wavelength "hot spots" at nanoscale edges and corners. Strong resonant optical absorption via Landau damping generates a large transient population of photo-excited electron-hole pairs at these hot spots in the metal[1]. The photo-excited carriers relax quickly, first through electron-electron scattering (~fs) to form a thermal distribution of "hot" electrons at a significantly elevated temperature compared to the lattice. On a slower timescale (~ps) electrons relax by scattering with phonons, resulting in local photo-thermal heating of the lattice[1]. There has recently been significant interest to take advantage of the high kinetic energy of these hot electrons, before they thermalize with the lattice, for applications in photodetection[2], optical energy conversion[3,4], and catalysis[5,6]. In addition to the short lifetime, one major challenge is the limited escape cone of hot electron trajectories with suitable momentum to exit the metal. Notably, nanoscopic confinement increases the probability that a hot electron will reach a surface with appropriate momentum for collection in an external device or chemical reaction[7,8].

In this article, we show how the extraordinary photophysics of plasmonic hot electrons are particularly well suited for optoelectronic power conversion based on thermionic power cycles. In a thermionic convertor, a metal cathode emitter is brought to high temperature so that a large fraction of the electrons has kinetic energy greater than the metal's work function. Electrons are vacuum emitted and collected at a lower temperature anode to generate electrical power. In principle, thermionic convertors have theoretical efficiencies far exceeding thermoelectric or mechanical energy converters[9]. In practice, the large energy barrier of the work function impedes electron emission unless temperatures exceed ~1500 K. This high temperature requirement has hindered practical application of the technology, which was explored historically for concentrated solar-thermal power generation[9,10], and more recently in conjunction with hybrid photovoltaic schemes [11,12]. However, the approximately 100-fold smaller heat capacity of the electron gas in metals[13], compared with the lattice, entails that nanostructured plasmonic absorbers can maintain a sub-population of hot electrons at a temperature greatly in excess of the lattice during steady-state illumination. Thus, optimized nanostructures can be used to effectively decouple the high kinetic energy electrons require for efficient thermionic emission while maintaining lattice temperatures compatible with the thermal tolerance of the metal.

In efforts to quantitatively understand photo-excited electron dynamics during thermionic power conversion, we have developed an optical thermometry technique based on anti-Stokes Raman spectroscopy that simultaneously determines lattice temperature, the hot electron temperature, and the size of the sub-population of hot electrons during steady-state illumination. Under optical powers spanning $10^6$-$10^{10}$ Wm$^{-2}$ on nanostructured gold we observe lattice temperatures between 300-600 K, while up to 8% of the electron gas is maintained at temperatures well above 1500 K. When these illuminated plasmonic absorbers are used as cathodes in thermionic power converters we indeed observe optical-to-electrical power conversion consistent with the electronic temperature and no evidence of thermal degradation. Moreover, we confirm this unique mechanism for decoupling electronic and lattice temperature overcomes challenges that have impeded other thermionic strategies for concentrated solar thermal power conversion. Further, our analysis outlines nanoscale design features that impart prolonged hot electron lifetimes, up to ~1 μs observed here, and other favorable electron dynamics that benefit broad classes of emerging hot electron technologies.

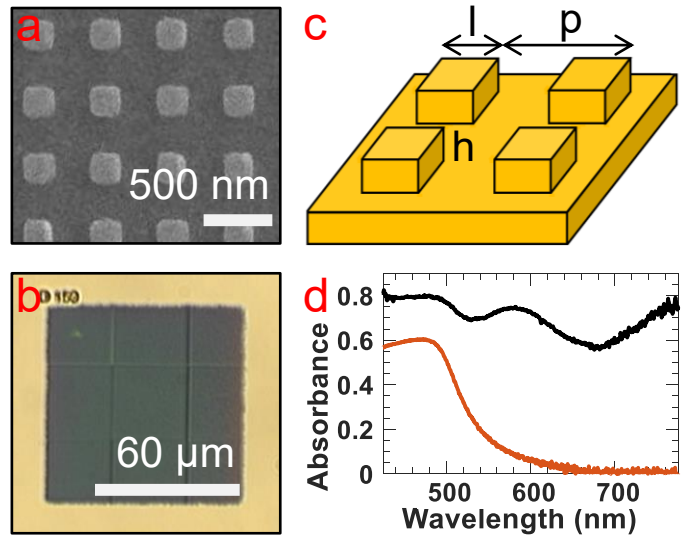

**Fig. 1.** (a) SEM and (b) optical image of the fabricated nanostructure. (c) Schematic of the unit cell with l =225 nm, p = 500 nm, and h = 100 nm on a 150 nm thick gold film. (d) The absorbance of the nanostructure (black) compared to a smooth gold thin film with thickness of 150 nm (red).

Top-down lithographic techniques were used to fabricate a 90 μm square array of 225 nm × 225 nm × 100 nm gold nanocubes at a pitch of 500 nm on a 150 nm thick gold film. The structure provides strong field concentration at hot spots on the nanocubes where there is greater surface area intersecting escape cones for hot electron emission[14] (See SI). Optical and SEM images are displayed in Fig. 1a, b. At the excitation wavelength (532 nm) there is an approximate 2-fold increase in absorption compared with a gold thin film (Fig. 1d), leading to increased photothermalization localized in the nanocubes.

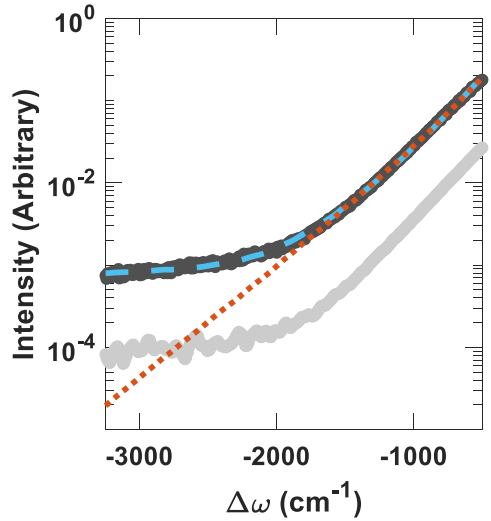

**Fig. 2.** Measured anti-Stokes Raman signal from the nanostructure (solid black) and gold thin film (solid grey), both collected under $7.3 \times 10^9$ Wm$^{-2}$ 532 nm laser excitation. The fit to a one-temperature model (eq. 1, red dotted) and two-temperature model (eq. 2, blue dash) are shown. The one-temperature fit gives $T_l = 459$ K, while the two-temperature fit gives $T_l = 430$ K, $T_e = 10,040$ K, and $\chi = 0.34\%$.

Temperature measurements during photothermalization were achieved by collecting anti-Stokes Raman spectra under 532 nm continuous wave (CW) laser illumination. A representative anti-Stokes spectrum is shown in Fig. 2. The signal from the nanostructure is ~10x larger than a gold thin film, comparable with enhancements observed in surface enhanced Raman studies[15]. Direct scattering from phonons does not contribute to the Raman signal from noble metals[16], therefore the broad frequency response is due to an anti-Stokes interaction directly with the electron gas. The physical origin of this anti-Stokes signal is still under some debate, with recent studies providing evidence that the signal may be due to photoluminescence rather than coherent scattering as in conventional Raman spectroscopy[17]. Regardless of the microscopic mechanism, the spectral-dependent intensity of the anti-Stokes spectrum has been established as an accurate indicator of the lattice temperature of a noble metal[18–21]. The anti-Stokes signal intensity therefore follows the Bose-Einstein thermal distribution of lattice excitations, as in eq. 1.

$$I(\Delta\omega) = C * D(\Delta\omega) * \left(\frac{1}{e^{\frac{hc\Delta\omega}{kT_l}} - 1}\right) \quad (1)$$

Here, I is the anti-Stokes signal intensity normalized by power and integration time as a function of the energy difference, $\Delta\omega$, from the Rayleigh line in m$^{-1}$, and $T_l$ is the lattice temperature in K. This expression includes a constant scaling factor, C, to account for experimental collection efficiency that is re-calibrated for each measurement. In addition, the signal intensity is proportional to the density of optical states, $D(\Delta\omega)$, obtained from the reflection spectrum.

Fitting our data to eq. 1 (Fig. 2, red dotted) proves inadequate as there is a large signal at high energy Raman shifts greater than -2000 cm$^{-1}$ that is not well described by the Bose-Einstein distribution. However, by adapting the method of Szczerbiński *et al.* our data is readily described if additional terms are included to account for a sub-population of hot electrons, χ, with an energy distribution at an elevated temperature, $T_e$[20].

$$I(\Delta\omega) = C * D(\Delta\omega) * \left(\frac{1-\chi}{e^{\frac{hc\Delta\omega}{kT_l}} - 1} + \frac{\chi}{e^{\frac{hc\Delta\omega}{kT_e}} + 1}\right) \quad (2)$$

The magnitude of χ in the steady state depends on both the generation rate of hot electrons due to optical excitation, and the relaxation rate as electrons equilibrate to $T_l$ via phonon scattering. Those carriers in thermal equilibrium with the lattice follow Bose-Einstein statistics, while the high-energy tail of the Raman signal is described by Fermi-Dirac statistics[22]. The fit to this two-temperature model (TTM) (Fig. 2, blue dash) is excellent for all samples and optical powers probed, spanning 10$^7$ to 10$^{11}$ Wm$^{-2}$. (See the SI for more details.)

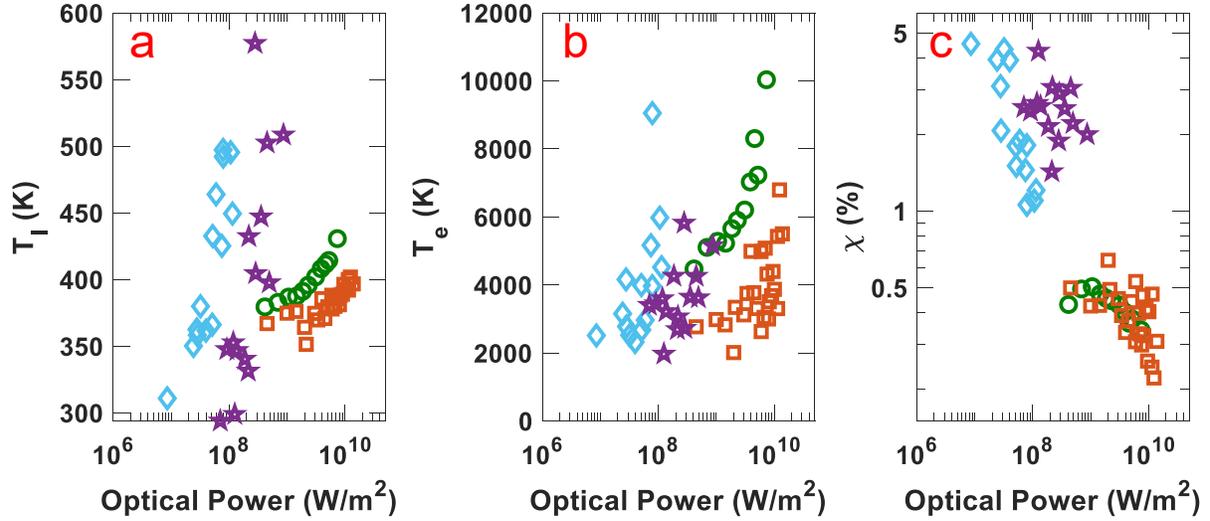

**Fig. 3.** The TTM fit for (a) lattice temperature, (b) electronic temperature, and (c) the percentage of hot electrons. These data are for the nanostructure under vacuum (blue diamonds), nanostructure in atmosphere (green circles), a gold thin film under vacuum (purple stars), and a gold thin film in atmosphere (red squares).

The dependence of $T_e$, $T_l$, and $\chi$ on optical power for the optimized nanostructure and for a 150 nm thick gold thin film control was determined by analyzing the Raman spectra using eq. 2. Samples were measured at atmosphere and under vacuum (0.010 mbar). The fitted data is summarized in Fig. 3. Melting and degradation of the samples occurred when the fitted $T_l$ significantly exceeded ~600 K in vacuum. In addition, the formation of a surface coating of gold oxide was apparent when $T_l$ exceeded ~450 K for samples in atmosphere (see SI Fig. S3), so data above those temperatures is omitted from this analysis[23]. In all experiments we observed a monotonic increase in $T_e$ and $T_l$ as the optical power increased, with $T_e$ in excess of $T_l$ by at least an order of magnitude. This trend is expected due to the lower heat capacity of the electron gas[13,24], and the values we measure for $T_e$ and $T_l$ are similar to those reported in transient absorption (TA) experiments[19]. To date, TA experiments have been the primary method for probing electron dynamics, however our experiments also access lower optical powers than can be achieved in pulsed time-resolved studies. Due to the decrease in convection, there is both a higher fitted $T_e$ and $T_l$ for samples under vacuum compared to samples in atmosphere. In all studies the nanostructure reaches significantly higher $T_e$ and $T_l$ than gold thin films at equivalent optical powers, due to the greater absorption and photothermalization.

A unique capability of our experiments that cannot be achieved readily in pulsed TA studies is quantification of the size of the hot electron population, $\chi$. An analysis of $\chi$ from our fitted spectra therefore provides important new information about how the availability of hot electrons depends on optical power and temperature under CW illumination that is more directly comparable to operating conditions for emerging hot-electron-based technologies. Interestingly in all experiments we observe a clear inverse correlation between $T_e$ and $\chi$ as optical power increases. One may initially expect that increases in optical power would lead to a greater rate of electronic excitation and thus a larger steady-state population of hot electrons. We hypothesize the opposite behavior is due to the increase in electron-phonon coupling as temperature increases, providing faster relaxation of the hot electrons that overwhelms the increase of the excitation rate. Our hypothesis is supported below by an analysis of the hot electron lifetime, $\tau$, and electron-phonon coupling constant, $G$, both calculated from $\chi$, also allowing for direct comparison of our findings with established TA measurements and computational studies[22,25].

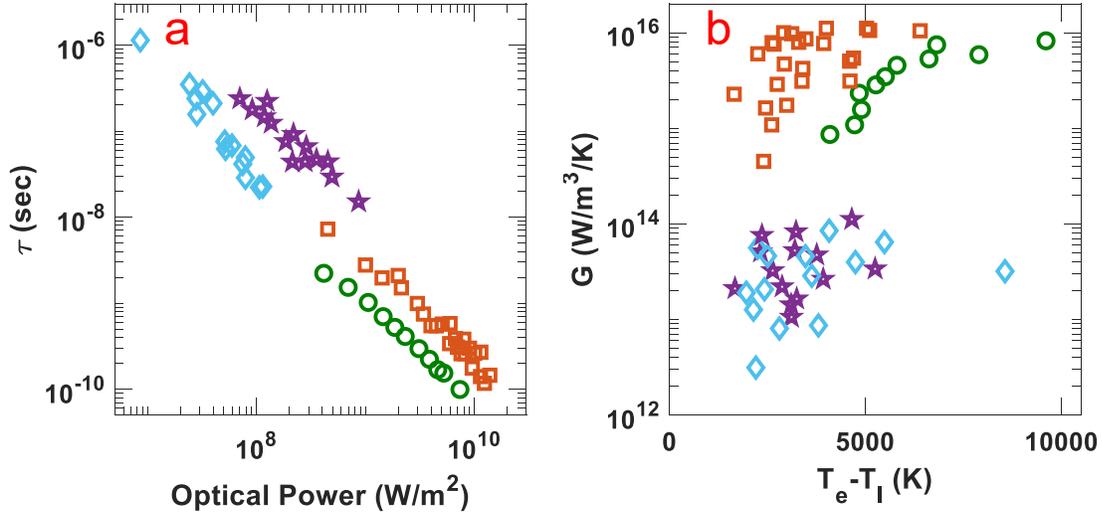

**Fig. 4.** (a) Calculated lifetime and (b) coupling constant for the nanostructure under vacuum (blue diamonds), nanostructure in atmosphere (green circles), gold thin film under vacuum (purple stars), and a gold thin film in atmosphere (red squares).

The lifetime of hot electrons within the elevated temperature distribution can be determined by comparing the size of the steady-state sub-population of hot electrons with the rate of hot electron generation. If it is assumed that every absorbed photon produces a transiently excited electron, then

$$\tau = \frac{\chi \rho V}{N \sigma} \quad (3)$$

where $\rho$ is the electron density of gold[26], V is the volume of the metal interacting with the light, N is the incident number of photons per second, and $\sigma$ is the experimentally measured absorbance. See the SI for more details. As can be seen in Fig. 4, for all four data sets there is a monotonic decrease in $\tau$ as the incident optical power is increased. At the highest optical powers $\tau$ approaches picosecond timescales, in agreement with TA measurements at similar powers[27]. Further, samples under vacuum show significantly longer $\tau$ than those at atmospheric pressure. We hypothesize that this difference may be due to surface collisions with gas molecules such as oxygen[23,28]. The observation of gold oxide formation at higher optical power provides further evidence that hot electrons interact with oxygen during illumination.

Further analysis of χ allows us to determine the electron-phonon coupling constant, G, independently from the lifetime. In the TTM well-established in TA studies, the time response of $T_e$ is related to volumetric electronic heat capacity, $C_e$, by the following relation[22]:

$$\chi \frac{\partial(C_e T_e)}{\partial t} = \chi G(T_e - T_l) + Q \quad (4)$$

Where G is the coupling constant in Wm$^{-3}$K$^{-1}$ and Q is the incident power in Wm$^{-3}$ coupled into the absorbing volume of the metal. We solve for G, as the time derivative goes to zero in the steady state. We have shown above that in atmosphere there are significant environmental contributions to the hot electron lifetime, implying that G accounts for coupling to all relaxation pathways. However in vacuum it is expected that electron-phonon coupling will dominate relaxation.

For all samples there is an increase in G as a function of temperature, in agreement with *ab initio* calculations and experimental studies[25]. In vacuum the environmental influences are minimized, and within the spread of the data, the nanostructure and thin film show an equivalent coupling constant that agrees with calculated values for nanoscale gold[22]. Notably, in atmosphere the gold thin film exhibits a larger G than the nanostructure at the same optical power. We hypothesize this trend in G is due to a decrease in the active surface area with hot electrons, likely near electromagnetic hot spots, and that only gas molecule collisions in these locations contribute to relaxation. The net result is that the nanostructure achieves much greater $T_e$ under equivalent optical power, and further, hot electrons have longer lifetimes compared with thin films at the same $T_e$ in atmosphere. Both behaviors are desirable in devices that take advantage of hot electrons, and our results suggest optical designs that decrease the relative volume in which hot electrons are generated could further optimize this response.

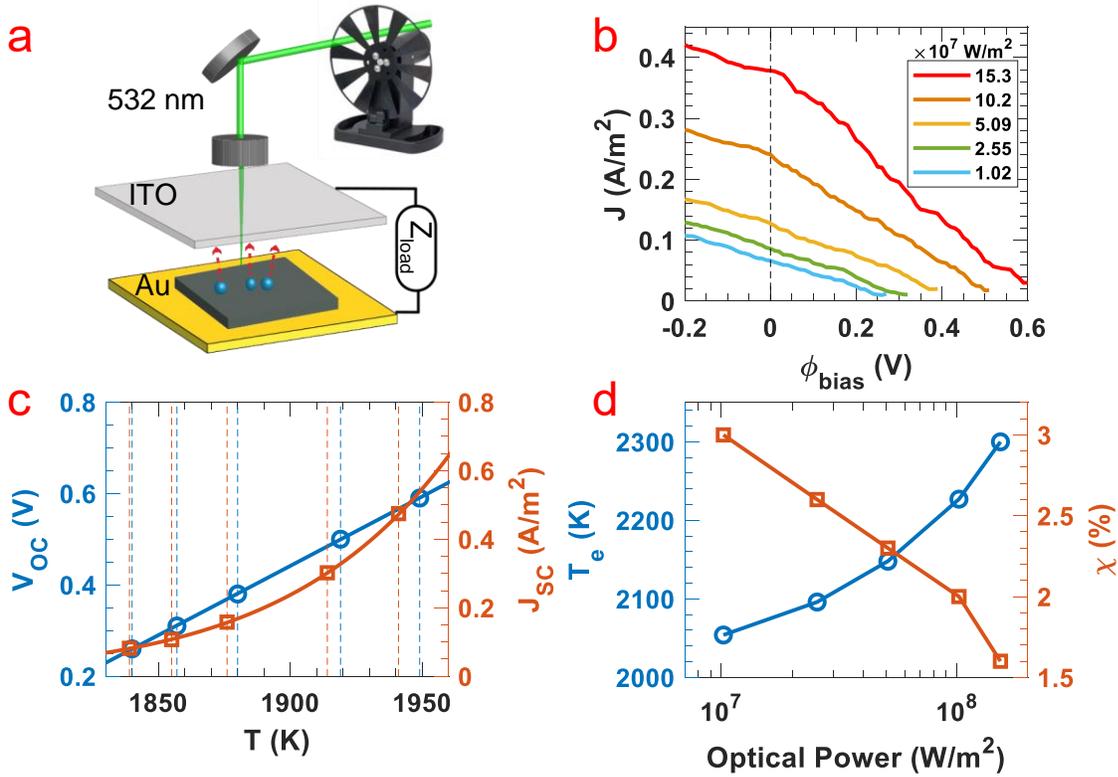

**Fig. 5.** (a) Schematic of thermionic emission measurement. (b) J-V curve measured at different optical powers. (c) Measured $J_{SC}$ (square) and $V_{OC}$ (circle) versus the calculated temperature according to a one-temperature model. The vertical dashed lines indicate the discrepancies in calculated temperature based on $J_{SC}$ (red) or $V_{OC}$ (blue). (d) Fitted electronic temperature (circle) and percentage of hot electrons χ (square) according to the two-temperature model of eq. 5.

In order to demonstrate that the hot electrons can perform work, we constructed a thermionic power convertor using the same nanostructure from Fig. 1 as an emitter with an ITO counter-electrode as a collector (Fig. 5a). The current density, J, was measured via a lock-in amplification scheme from parallel electrodes separated by 200 μm during 532 nm CW illumination under vacuum (0.010 mbar), see SI for further details. The power generation region of the current-voltage (J-V) response is depicted in Fig. 5b. The open-circuit voltage ($V_{OC}$) reported here represents the retarding bias at which the current density reaches the noise level for the lock-in amplifier (see SI). The downward curvature of the J-V response indicates the presence of space charge effects during the measurement[9].

Thermionic emission current density is conventionally described using Richardson's equation, which we adapted to accommodate a TTM. Because only the fraction of hot electrons, $\chi$, at temperature $T_e$ provide a non-negligible contribution to the thermionic current,

$$J = \chi A T_e^2 e^{\frac{-(W+\phi_{bias}+\phi_{sc})}{kT_e}} \quad (5)$$

where $A = 1.2017 \times 10^7$ Am$^{-2}$K$^{-2}$ is the Richardson's constant, $W = 5.1$ eV is the work function of gold[29], $\phi_{bias}$ is the external potential with the positive sign indicating a retarding bias, and $\phi_{SC}$ is the additional potential due to the electrostatic field of the space charge, calculated using Langmuir's space charge theory[30]. If we analyze our data assuming a one temperature model, then the short circuit current at zero bias, $J_{SC}$, and the $V_{OC}$ measured from the J-V response are inconsistent with a unique fitted temperature. However, for each optical power probed there is a unique combination of $T_e$ and $\chi$ that can be input into eq. 5 to accurately reproduce both the experimentally measured $J_{SC}$ and $V_{OC}$, as summarized in Fig. 5d.

We find that $T_e$ monotonically increased with optical power, with the same inverse relationship between $\chi$ and $T_e$ measured in the Raman studies above. Further, the magnitude of $\chi$ is consistent with the Raman fitting under the same optical powers, though the fitted $T_e$ based on the device response is somewhat lower. Several factors could give rise to this discrepancy, and we hypothesize that the largest source of error may be due to the more complex geometry of the space charge for electrons emitted by the nanostructure surface in comparison with the parallel plate geometry assumed in Langmuir's space charge theory[30]. (See SI for more discussion.) We note that the inverse correlation between $T_e$ and $\chi$ is manifest independent of what analysis is used to model the space charge effect or other non-idealities not accounted for in eq. 5. As an important point of comparison, we observed no measurable current under any optical power when a gold thin film was used as the emitter, even though both the thin film and nanostructure reached comparable $T_e$ according to the Raman fitting. This difference is likely due to a 3-fold increase, at minimum, in suitable escape cones for hot electrons in the nanocubes, and highlights how momentum constraints are relaxed in the plasmonic nanostructure.

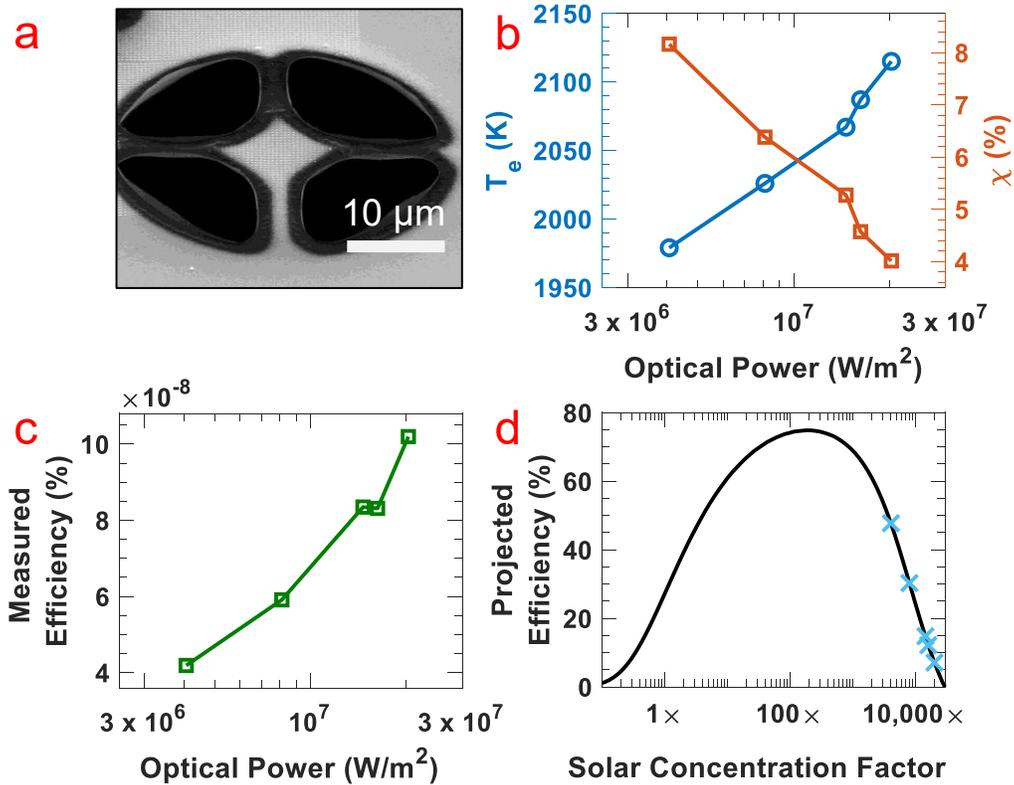

**Fig. 6.** (a) SEM image of a thermally isolated gold nanostructure (b) Fitted electronic temperature (circles) and percentage of hot electrons (squares) according to the two-temperature model of eq. 5 (c) Measured optical power conversion efficiency and (d) projected optical power conversion as a function of solar concentration factor, assuming a decreased work function, W = 1.6 eV. The blue crosses correspond to the optical powers probed experimentally.

To demonstrate the potential of this strategy for solar power conversion, an additional sample was prepared that minimized losses due to conduction. The nanostructure was fabricated on a 50 nm thick $Si_3N_4$ membrane. Focused ion beam etching was used to perforate the membrane and thermally isolate a 6 x 6 μm section of the array (Fig. 6). In vacuum, the device achieved optical power conversion efficiency between $10^{-8} – 10^{-7}$ %, under $4 \times 10^6 – 2.1 \times 10^7$ $Wm^{-2}$. This optical power range is comparable to that employed in solar-thermal conversion schemes, where solar concentration factors are commonly between 1500 – 4000x. While the sample showed no evidence of thermal degradation, the seemingly low efficiency is due to the large work function of gold, W = 5.1 eV. It is common practice during thermionic device operation to include rarified Cs metal vapor to both decrease W via surface adsorption and minimize space charge effects. Gold surfaces with sub-monolayer cesium have a reported work function of W = 1.6 eV[31]. Assuming the same photo-thermal response measured here but with W = 1.6 eV, a maximum conversion efficiency of 74.9% is predicted to occur at 190x solar concentration, based on the trade-off between $T_e$ and $\chi$. If practically achievable, such high efficiency for collecting hot electrons would significantly decrease the optical energy that is available to promote heating of the lattice through electron-phonon coupling, further promoting stability of the emitter. See the SI for more details on this calculation. For comparison, state-of-the-art solar-thermal conversion strategies achieve ~30% efficiency commonly at temperatures greater than 1000 K [32].

In conclusion, we have demonstrated a new opto-electronic power conversion mechanism that uses plasmonic nanostructures to decouple electronic temperature and lattice temperature during steady-state optical illumination of a thermionic emitter. We have also developed an optical thermometry technique based on anti-Stokes Raman spectroscopy to quantify these separate temperatures, as well as the size of the sub-population of hot electrons. Our results show an inverse relationship between the temperature and the population of the hot electron gas, and analysis of the lifetime and electron-phonon coupling show how designs that decrease the volume of the metal can further optimize the hot electron dynamics. When integrated into thermionic devices the plasmonic cathodes provide optical power conversion efficiency consistent with the electronic temperature, while maintaining significantly lower lattice temperatures. Thus, we demonstrated how this mechanism can overcome challenges related to thermal stability that have historically limited the use of thermionic devices for solar-thermal energy conversion. We believe the remarkable tailorability of plasmonic nanostructures may allow further opportunities for very efficient solar energy conversion based on this strategy.

**Methods**

FABRICATION: To prepare the nanostructures, a 5 nm Cr sticking layer followed by a 150 nm Au thin film were deposited using thermal deposition (Lesker PVD e-beam evaporator) on a commercially available silicon TEM grid with a 50 nm $Si_3N_4$ membrane windows (Ted Pella). A layer of PMMA/MMA 9% in ethyl lactate (MicroChem) followed by a layer of 2% PMMA in anisole (MicroChem) were spin coated to form a bilayer resist. Electron beam lithography was performed using a Tescan FE-SEM instrument. A top Au layer was deposited followed by removal of the polymer mask in acetone. For thermally isolated samples, focused ion beam (FIB) etching was performed using a Xe source. (Tescan FERA-3 FIB-SEM).

SPECTROSCOPY: Anti-Stokes Raman spectra were collected using a confocal Raman microscope (Witec RA300) with samples in a vacuum heating microscope stage attached to a vacuum pump (Linkam TS1500VE). Vacuum experiments were performed at a pressure of 0.011 mbar. Samples were illuminated by a Nd:Yag laser at 532 nm and focused on the sample using a 20x objective with a 0.4 NA. Reflection spectra were taken by using the same stage setup with a white light source. The measured reflection signal was normalized to the source spectrum to give the reflectance of the surfaces.

DEVICE: For the measurement of the hot electron thermionic emission current density, a pair of parallel electrodes composed of nanostructured gold patterns and an ITO glass slide was constructed. A 200 μm spacer separating the two electrodes was made of Kapton tape (attached to ITO glass) and copper tape (attached to the substrate of gold nanostructures) to ensure good electrical contact. The assembled electrodes were placed in the same microscope stage as above (Linkam TS1500VE) to measure the thermionic emission current under vacuum ($<10^{-5}$ Torr). The electrodes were connected to a source-measure unit (Keithley 2450) in order to measure the current at varying bias voltage. The light from a CW diode laser emitting at 532 nm was used for the photoexcitation, which was focused on the gold nanostructures through a 50× objective to a focus spot on the sample with a beam diameter of ~5.6 μm. The thermionic emission current was measured with a lock-in amplifier (Stanford Research Systems, SR830) by chopping the excitation light at 47 Hz. For each power density used, the bias voltage was swept from -0.2 V (accelerating bias) to 1 V (retarding bias).

## **Acknowledgements**

This work is funded by the Air Force Office of Scientific Research under award number FA9550-16-1-0154. M.S. also acknowledges support from the Welch Foundation (A-1886).

## **Author Contributions**

N.H. fabricated devices and performed spectroscopy. S.W performed electronic device measurements. M.S. supervised the project. All authors participated in data analysis.

**Competing Interests**

Authors declare no competing interests.

# Hot Electron Dynamics in Plasmonic Thermionic Emitters

Nicki Hogan[1], Shengxiang Wu[1], Matthew Sheldon*[1,2]

[1]Department of Chemistry, Texas A&M University, College Station, TX, 77843-3255, USA.

[2]Department of Material Science and Engineering, Texas A&M University, College Station, TX, 77843-3255, USA.

*sheldonm@tamu.edu

**Supplemental Information**

Field Enhancement Map of Nanocubes with 532 nm Excitation

To demonstrate the location of electromagnetic 'hot spots' in the gold nanostructure, we calculated the electromagnetic field enhancement using full wave optical simulations (FDTD method, Lumerical Inc.). The nanostructure simulated was a 225 x 225 x100 nm gold cube on a 150 nm gold film, with the Au refractive index from Johnson and Christy[33]. Light was injected using a planewave source and monitored at 532 nm. Periodic boundary conditions were used to simulate an infinite array with periodicity equivalent to the fabricated samples analyzed in the study. Maximum field enhancement is localized to edges and corners of the nanocubes with a maximum field enhancement of approximately 185x at the corners.

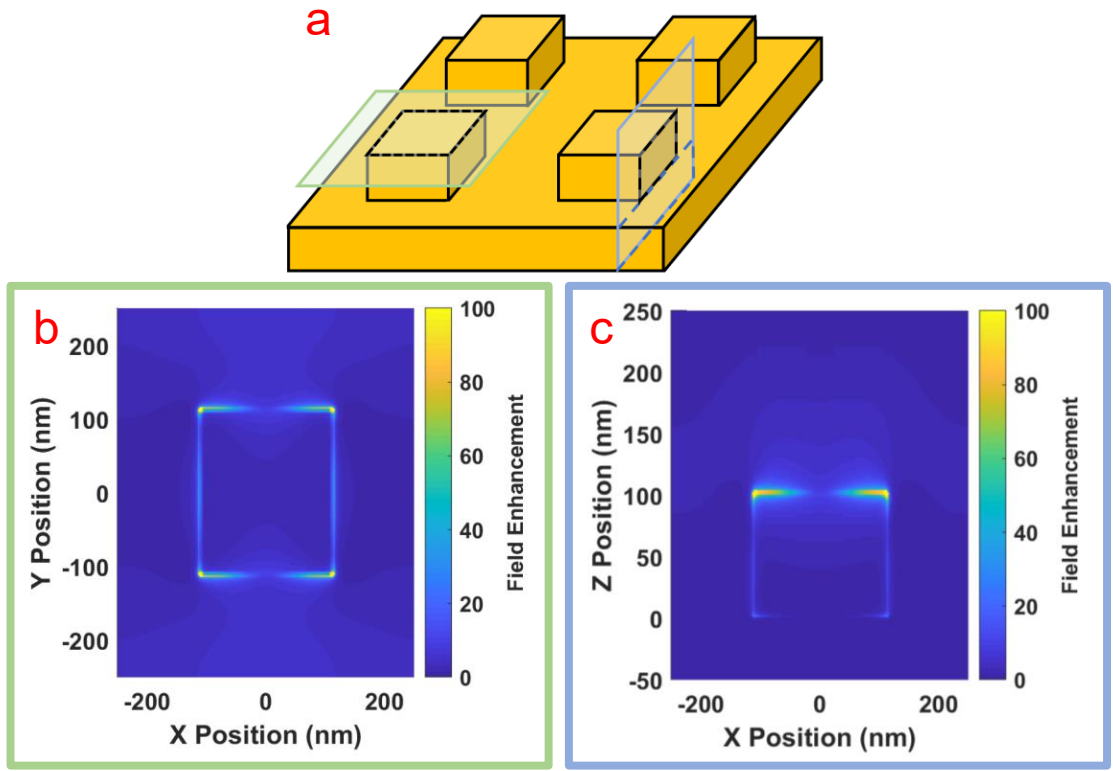

**Fig. S1.** (a) Schematic of the locations monitored on the surface of the nanostructure with the corresponding field enhancement along (b) the top face of the nanocube and (c) an edge of one nanocube.

Representative Full Raman Spectrum

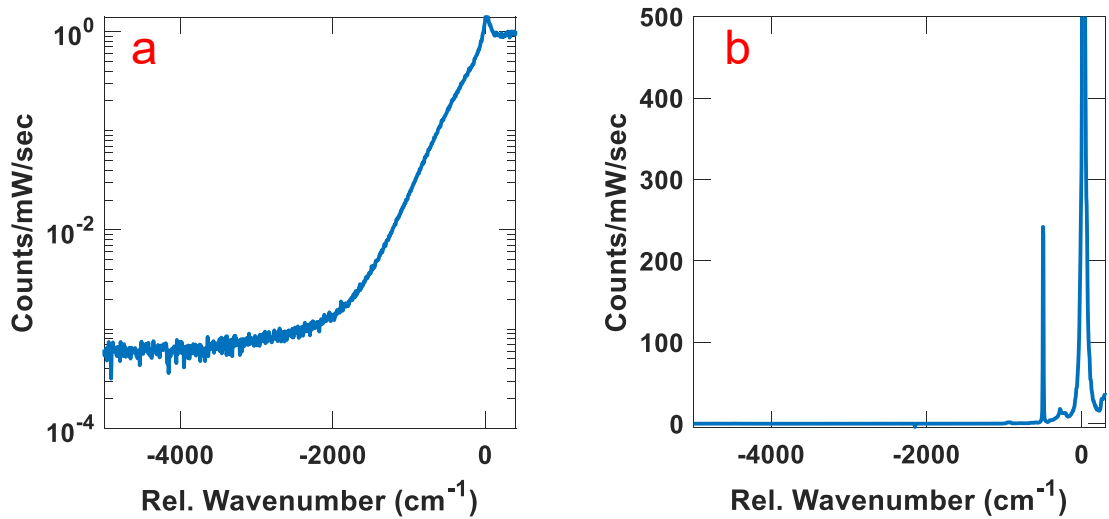

**Fig. S2.** (a) A representative anti-Stokes spectrum across the entire spectral range collected during the measurement. Data in the spectral range between -3500 – -500 cm$^{-1}$ was fitted according eq. 2 in the main text. This spectrum corresponds to a gold nanostructure in atmosphere under an optical power of 4.6 x 10$^9$ W/m$^2$. The peak at 0 cm$^{-1}$ is due to the Rayleigh line from laser excitation not blocked by the filter. The width of the laser line can be seen in (b) which is the spectra of a silicon substrate at the same incident power. The laser line tails off at around -500 cm$^{-1}$ so that spectral range was omitted during fitting.

Evidence of Gold Oxide

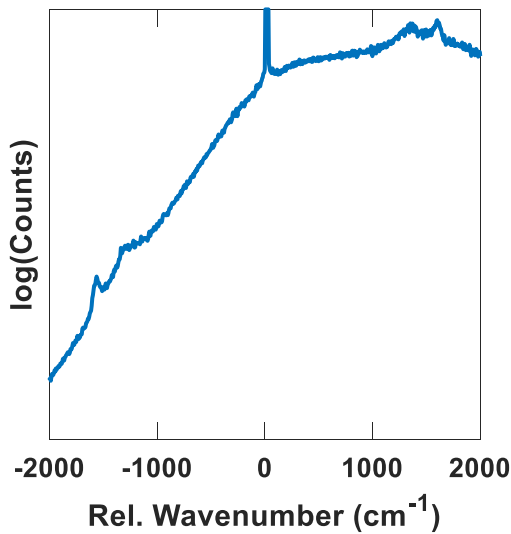

**Fig. S3.** A representative spectra from a gold film at 1 x $10^{10}$ W/m$^2$ which shows the formation of Raman peaks at 1350 and 1580 cm$^{-1}$ and their corresponding anti-Stokes peaks. We attribute these peaks to the formation of surface gold oxide under high optical power.

Fitting the Anti-Stokes Raman Signal to the Two-Temperature Model

The Anti-stokes emission data is fit to eq. 2 using a linear least squares analysis with four free fit parameters: $T_l, T_e, \chi$, and C. The quality of fit was determined by minimizing the squares of the residuals. In order to prevent finding local minima in the residual space we fit the data by systematically varying all combinations of starting conditions for the fitting routine across 5 orders of magnitude, and then found the solution that gave the global minimum of the residual. In addition, due to the relatively smaller magnitude of the signal at higher energies, where the electronic temperature is the dominant contribution to the signal, we instead fit to the log$^2$ of the data, in order to put greater weight on the component of the signal arising from the hot electron temperature.

Using a Boltzmann Distribution Instead of a Fermi-Dirac Distribution

In addition to using Fermi-Dirac statistics to model the hot electron distribution, we also fitted for temperature assuming the contribution from the hot electron temperature could be modeled as a Boltzmann distribution. Then, eq. 2 can be rewritten as the following:

$$I(\Delta\omega) = C * D(\Delta\omega) * \left( \frac{1-\chi}{e^{\frac{hc\Delta\omega}{kT_l}} - 1} + \frac{\chi}{e^{\frac{hc\Delta\omega}{kT_e}}} \right) \quad (S1)$$

The Raman spectra fit to this expression with equal fidelity as the data reported in the main manuscript, however the fitted electronic temperatures are significantly higher than what is reported in other literature at equivalent optical power[17,19,20].

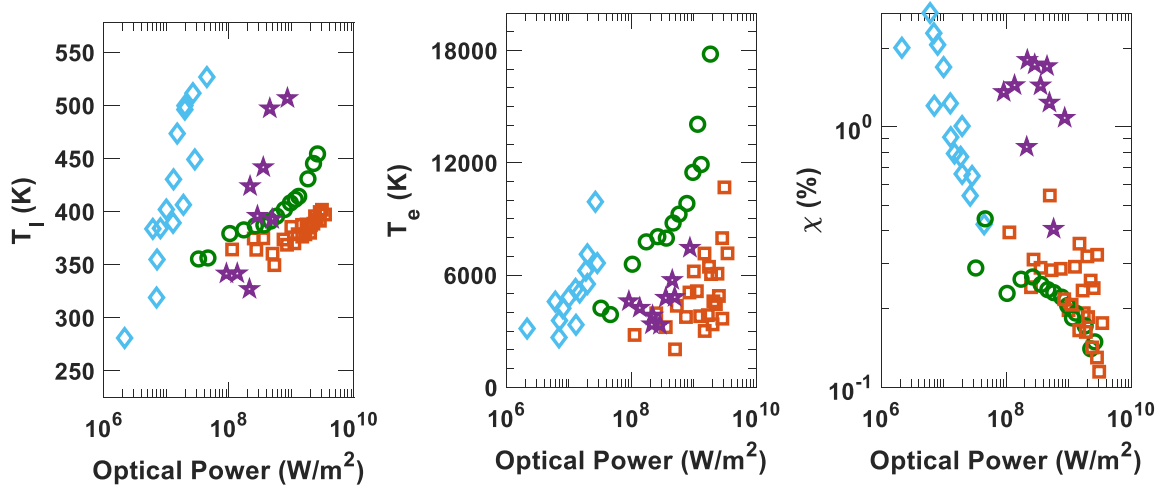

**Fig. S4.** (a) The lattice and (b) electronic temperature as well as (c) the fraction of hot electrons when using Boltzmann statistics to fit the high-energy region of the spectrum. These data are for the nanostructure under vacuum (blue diamonds), nanostructure in atmosphere (green circles), a gold thin film in atmosphere (red squares), and a gold thin film under vacuum (purple stars).

Calculating Interaction Volume

The interaction volume of the excitation source at the sample surface for the calculation of the lifetime in eq. 3 is determined using the spot size of the laser on the surface and the penetration depth of the light into the material. The spot size was determined using an optical image to have a radius of 0.8 µm. The penetration depth was calculated from the absorption coefficient[34]. The absorption coefficient, α, is a function of the imaginary component of the refractive index, n, at the wavelength used for excitation this study (532 nm). Based on the dielectric function of Au reported by Johnson and Christy[33].

$$\alpha = \frac{4\pi}{\lambda} \text{Im}(n) = 5.26 \times 10^5 \text{ cm}^{-1} \quad \quad (S2)$$

The penetration depth is defined as the distance at which light has decayed to 1/e intensity compared to the incident intensity[34].

$$d = \frac{1}{\alpha} = 19 \text{ nm} \quad \quad (S3)$$

Determination of $V_{OC}$ During Device Measurements

As shown in Fig S5, we define open circuit voltage, $V_{OC}$, as the bias voltage at which the lock-in amplifier can no longer lock onto the signal. The ITO does not contribute a reverse current during the experiment.

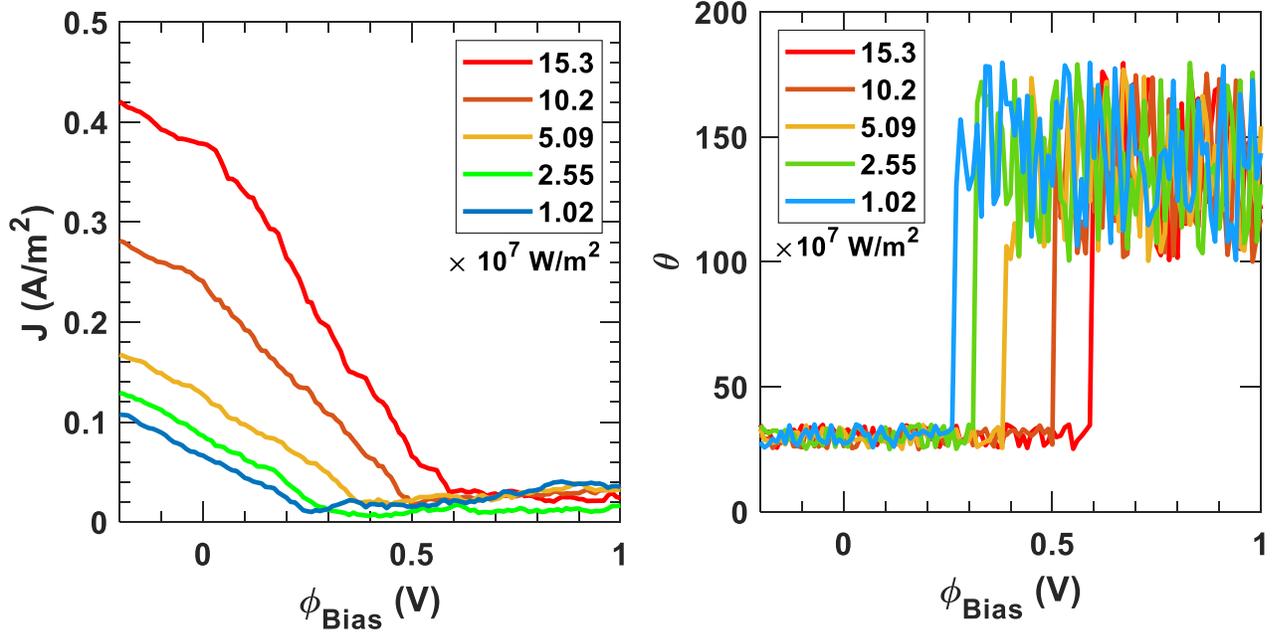

**Fig. S5.** a) J-V curve measured at different power densities and (b) Corresponding lock-in phase vs applying bias voltage recorded at each power density.

Determination of $T_e$ Using the Modified Richardson's Equation

$$J = \chi A T_e^2 e^{\frac{-(W+\phi_{bias}+\phi_{SC})}{kT_e}} \tag{S4}$$

Assuming the hot carriers at an elevated electronic temperature comprise a small fraction of the total electron gas during steady state illumination, we determined a unique combination of $T_e$ and $\chi$ graphically, by plotting the iso-lines of $J_{SC}$ and $V_{OC}$ at each incident optical power, as shown in panels b-f in Fig. S6. Each iso-line corresponds to any combination of $T_e$ and $\chi$ that gives the measured value of $J_{SC}$ (blue) or $V_{OC}$ (red) at that optical power. We see that there is only one combination of $T_e$ and $\chi$ that is consistent with both measured quantities.

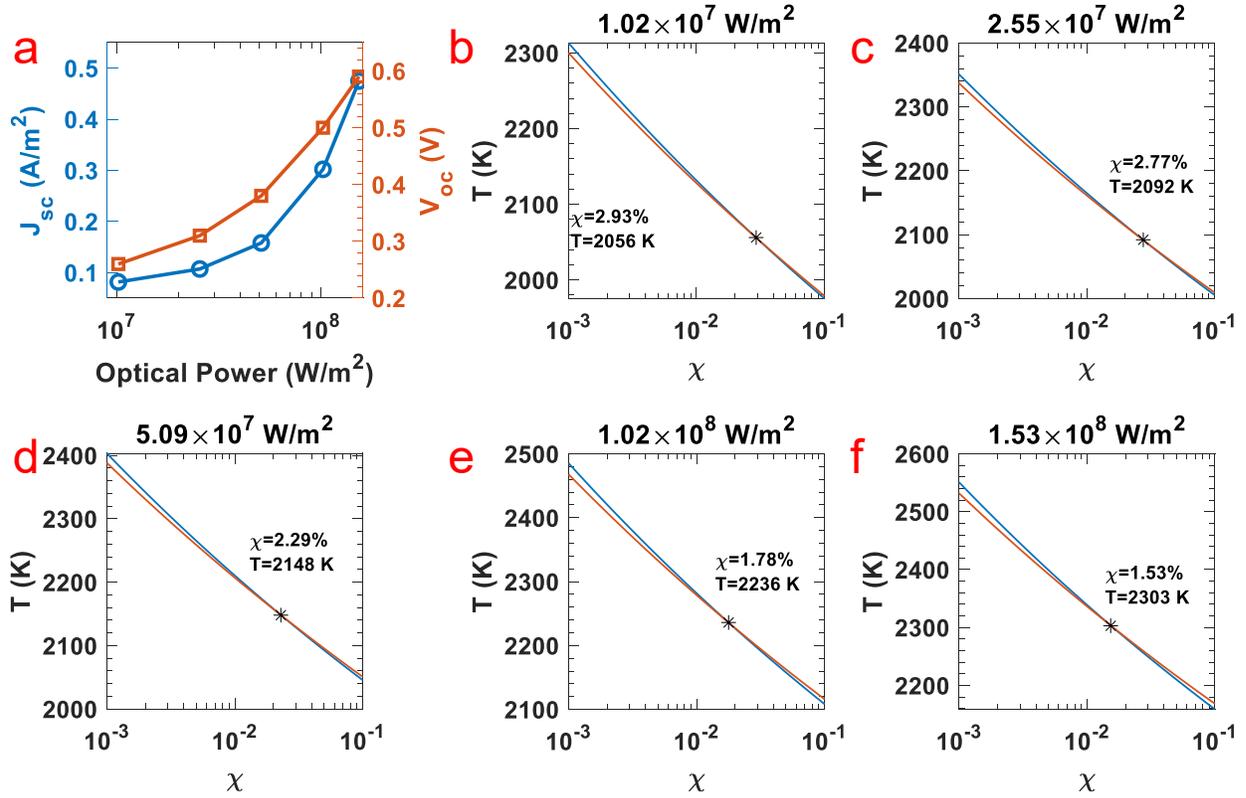

**Fig. S6.** (a) Experimentally determined $J_{SC}$ and $V_{OC}$ vs. optical power, (b)-(f) Determination of electronic temperature and hot electron fraction using the modified Richardson's equation (blue lines and red lines are iso-lines of $J_{SC}$ and $V_{OC}$ respectively).

Space Charge Potential

The space charge potential is calculated using Child-Langmuir theory across a vacuum junction with a gap distance, d, in which the saturation thermionic emission current $J_{ES}$ is first calculated using Richardson's equation at a given $T_e$ and $\chi$. The critical current $J_R$ is then calculated using eq. S5(*30*).

$$J_R = 9.664 \times 10^{-6} * \frac{\left(\frac{kT}{e}\right)^{3/2}}{d^2} \tag{S5}$$

The experimentally determined currents reside in the range between $J_R$ and $J_{ES}$ indicating that space charge limited the current collected during experiments. Therefore, the additional space charge potential is calculated using eq. S6[30].

$$\phi_{sc} = ekT_e \ln\frac{J_{ES}}{J} \qquad (S6)$$

Where e is electron charge, k is Boltzmann constant.

Summary of Measurements from an Isolated Nanostructure

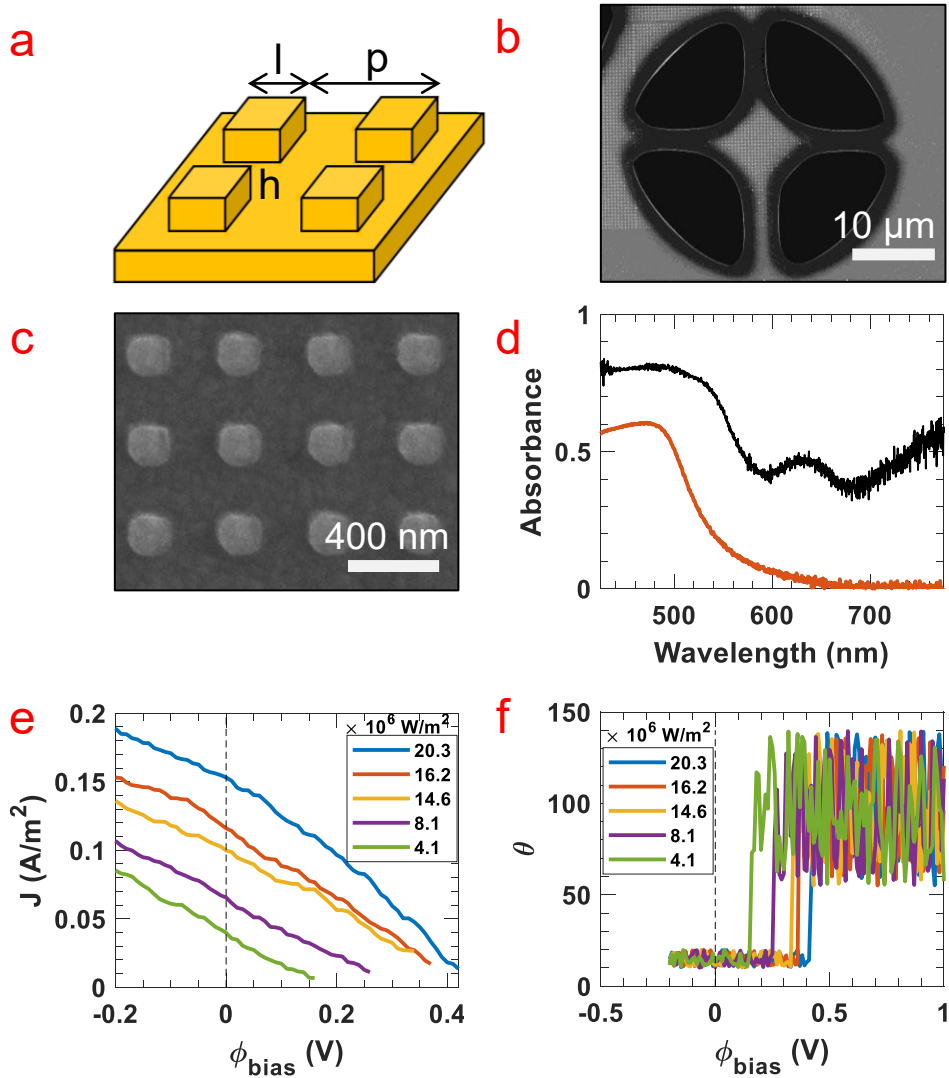

**Fig. S7.** (a) Schematic of gold nanostructure, (l = 160 nm, p = 380 nm, h = 100 nm)(b), (c) SEM images of thermally-isolated gold nanostructure, (d) Absorption of gold nanostructures (black) and gold thin film (red), (e) J-V response of thermally-isolated gold nanostructures and (f) Corresponding phase dependence during lock-in measurement.

Projected Efficiency for Thermally Isolated Gold Nanostructure with Cs

Due to the high work function of Au (W = 5.1 eV), there is a low thermionic current at the electronic temperatures induced during our experiments. At low current density, the emitted electrons do not remove enough energy from the system to significantly perturb the electronic and lattice temperature, similar to a theoretical scenario in which no current is emitted and all optical power goes to photo-thermalization. Thus, the J-V response we measured provides a calibration that relates both the electronic temperature and the population of hot electrons at a given incident optical power (SI Fig. S8, blue data), when the temperature of the system is not lowered by the collection of thermionic current. We extrapolate this relationship to zero incident optical power to ensure 100% of the electrons are at room temperature (SI Fig. S8, red trace).

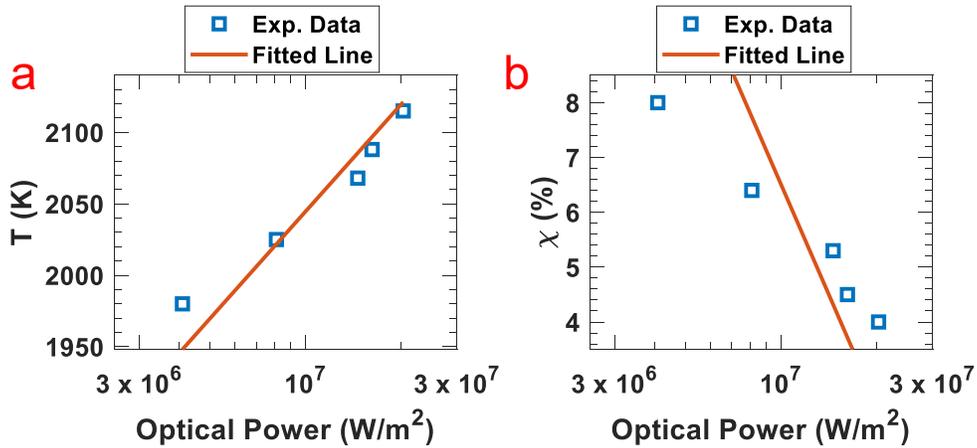

**Fig. S8.** (a) Experimentally determined $T_e$ vs optical power and the fitted relationship (red trace), ensuring zero incident power gives $T_e = 300$ K (b) Experimentally determined $\chi$ vs optical power and fitted relationship (red trace) ensuring that zero incident optical power gives $\chi = 100\%$.

However, if the work function of the cathode is lowered by the addition of Cs, the increase in emission current can draw enough electrical power, $P_{electrical}$, from the system to significantly lower the steady-state electronic and lattice temperature. Therefore, the fraction of the absorbed power $P_{heating}$ that can contribute to heating is decreased compared with the incident optical power, $P_{optical}$:

$$P_{heating} = P_{optical} - P_{electrical}$$

$$\swarrow \qquad \searrow$$

$$T_e, \chi$$

By these relations, $P_{electrical}$ can be calculated assuming a given $T_e$ and $\chi$, which is further constrained by the power $P_{heating}$ available to heat the electron gas. Thus, this imposes a self-consistent condition that results in a unique combination of $T_e$ and $\chi$ at any incident optical power, as well as the corresponding $P_{electrical}$, based on the calibration above. The summary of this result is the data in Fig. 6 in the main manuscript, reproduced in SI Fig. S9 below.

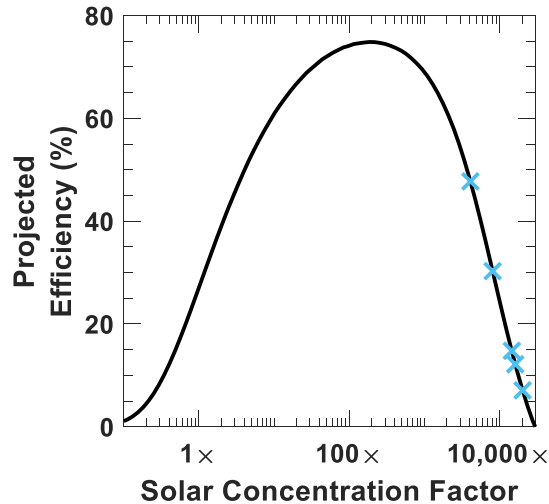

**Fig. S9.** The projected efficiency of an isolated gold nanostructure thermionic emitter as a function of solar concentration factor, assuming the device is in an atmosphere of rarified Cs vapor with the gold work function W = 1.6 eV. The blue crosses correspond to the optical powers measured experimentally.

**Additional References**